\documentclass[conference]{IEEEtran}
\ifCLASSINFOpdf
   \usepackage[pdftex]{graphicx}
\else
\fi
\usepackage{url}

\begin{document}
%
\title{Topological Filtering for Visual Data Mining and Analysis of Complex Networks}



\author{\IEEEauthorblockN{Faraz Zaidi}
\IEEEauthorblockA{CNRS UMR 5800 LaBRI \& INRIA Bordeaux - Sud Ouest\\ 
    				351, cours de la Lib\'{e}ration\\ 
    				33405 Talence cedex, FRANCE \\           
					Email: faraz.zaidi@labri.fr}
}


%


\maketitle

\begin{abstract}

The discovery of small world and scale free properties of many real world networks has revolutionized the way we study, analyze, model and process networks. An important way to analyze these complex networks is to visualize them using graph layout algorithms. Due to their large size and complex connectivity, it is  difficult to make deductions from the visual representation of these networks. In this paper, we present a method for interactive analysis of large graphs based on topological filtering, network metrics and visualization. We analyze a number of real world networks and draw interesting conclusions using the proposed method.


\end{abstract}


%
\IEEEpeerreviewmaketitle

\section{Introduction}

Recent years have seen the ability of computers to store, process and visualize large size networks~\cite{landesberger11}. The term \textit{network} has different significations for people from different fields and is used extensively to represent systems such as social networks~\cite{wasserman94,bourqui09,bourqui08}, online social media~\cite{comar12}, web graphs~\cite{zaidi09} networks~\cite{bourqui09,bourqui08},  air transport~\cite{rozenblat08}, metabolic pathways~\cite{bilke01} and world wide web~\cite{adamic99}. Research has shown that although seemingly diverse, these fields have strong common methodological foundations and share methods to analyze, model, understand and organize these networks. Different measures have been studied to classify these networks~\cite{brandes05}. Two such classifications have gained lots of interest when networks exhibit \textit{small world}~\cite{watts98} and \textit{scale free}~\cite{barabasi99} features. 


Having a look at the taxonomy of existing methods to analyze networks, one way is to develop metrics and measures that return quantitative knowledge about these networks. These measures can be grouped as \textit{element level}, \textit{group level} and \textit{network level}~\cite{brandes05}. A good example of an element level measure is the degree of a node, which refers to the number of connections a node has to other nodes in the network. On the other hand, network level metrics consider the entire network to calculate a quantitative measure. Betweeness Centrality~\cite{freeman77} is a good example of a network level metric.

Another way is to use visualization to understand and comprehend these networks. These methods can be grouped into three classes as \textit{visual exploration of data}, \textit{visualization of data mining results} and  \textit{tightly coupled data mining and visualization}~\cite{oliveira03, zaidi10}. Visual data exploration is the interactive exploration of graphical representation of data to discover knowledge. Visualization of data mining results is the process of graphically representing the results of a data mining process to enable humans to easily interpret the obtained results. A hybrid form of these two methods is where visualization is integrated as part of the data mining algorithm and users or domain experts interactively guide the data mining algorithm. Visual exploration is a useful method to discover hidden knowledge and extract interesting patterns in data~\cite{keim02}. It has been effectively applied in a number of different fields with graphs following small world and scale free properties~\cite{Jia11}. 

Most of the real world networks today have hundreds of thousands of nodes and edges. Metrics that are computationally expensive are no longer practically applicable for real world networks. Even with the increasing computational power, we still need faster methods, heuristics and decomposition methods that can process these networks in reasonable time frames. Moreover, for these large size networks, measures like centrality and structural organization of networks are no longer dominated by elements but rather, groups and subsets. The complex interactivity of these subsets play an important role in the overall behavior and evolution of networks, thus presenting researchers with a challenging problem.

An interesting approach to solve this problem is to decompose a network into sub-components or sub-networks and then apply either measures or visualization methods on these sub-networks to gain insight of the whole networks. In terms of a measure, the decomposition speeds up the calculation as a subset of nodes is considered instead of the entire network. In terms of visualization, it is easier to visually analyze a network with few nodes as compared to the entire network. Decomposition techniques are significant for visualization methods as it is practically impossible to visualize these large size networks on a computer screen even with the new and advanced visualization tools and interactive techniques for visualization~\cite{herman00}.


The contribution of this work is not the results or findings that we obtained about different networks. All the findings listed in the paper are well known (and they are cited) but these were discovered through several different methods, often requiring complex mathematical and statistical formulations and derivations. The contribution of this work is the ease of the proposed method, as domain experts from a variety of different fields can benefit without requiring any special skills. The proposed method is general and applicable to a variety of networks as shown in the case studies when applied to social, biological and technological networks. Furthermore, network metrics, filtering and visualization methods have never been combined in this way to discover these findings. Our objective is to rediscover the known facts and findings using another approach which is suited for domain experts with non-mathematical and non-computer science background to understand the complex interconnected behavior of many different real world networks.

In this paper, we introduce a visual data exploration method which decomposes large graphs through topology based node filtering, uses networks metrics for quantitative analysis and visualization for macro level analysis and interactive exploration. This topological decomposition is motivated by two important features of real networks. The first is that these networks have power law degree distribution and the second, visualization of these networks produce highly entangled and hard to read drawings~\cite{henry07}. Our idea is to create sub-networks based on degree connectivity and then visualize these sub-networks to understand, analyze and extract information from them. 


The rest of the paper is organized as follows: We introduce a number of networks used for experimentation throughout this article in section~\ref{sec::data}. In section~\ref{sec::data}, we introduce the proposed method and explain how various data sets can be analyzed using this method. Finally we present conclusions and future research directions in section~\ref{sec::conclusions}.

\section{Related Work}

The idea to decompose and study complex networks is not new. A decomposition based on the connectivity of vertices was proposed by Batagelj and Zaversnik called the $k$-core decomposition~\cite{batagelj02}. The method consists of identifying subsets of the network called $k$-cores. These subsets are obtained by recursively removing all the vertices of degree smaller than $k$, until the degree of all remaining vertices is larger than or equal to $k$. So for example, to obtain a $2-core$ of a network, we remove all the vertices with degree less than $2$ in the network, which is nodes of degree $1$. After this removal, certain vertices that previously had high degree might now have degree $1$, the removal process is repeated again until there are no vertices of degree $1$ left in the network. All the vertices left after this removal become part of $2-core$ and the process can be restarted for higher values of $k$.

Cores with larger values of $k$ correspond to sets of vertices with high degree and connected to high degree vertices only. This gives cores with larger values of $k$, a more central position in the network's structure~\cite{hamelin06}. This method has been used in several domains to analyze networks and the connectivity of vertices such as, in the analysis of protein interaction networks~\cite{wuchty05}. Apart from its utilization in analysis, it has also been used to visualize large scale networks as it decomposes a network into subsets of vertices of increasing centrality. It can also help focus on certain regions of interest in a network~\cite{hamelin06}.

The method we propose is significantly different from $k$-cores, although both $k$-core and the proposed topological decomposition are based on degree of vertices and create subsets. Topological decomposition focuses on studying how edges are distributed in high and low degree nodes. $k$-cores focus on recursively identifying central nodes and has clearly different objectives. The differences will become more evident as we explain the details of our method.

There are a number of other ways to decompose a network. The most widely used method is \textit{Clustering}~\cite{schaeffer07}, which groups similar nodes together. Clustering has often been used to decompose large networks for better understanding and visual mining \cite{keim02}. Clustering focuses on grouping similar nodes whereas our objective is to decompose the network and simplify the complex connectivity behavior of nodes. The toplogical similarity of nodes can be used as a similarity metric to group nodes together and thus the proposed method can be considered as a topology based clustering method. For simplicity, we refer to our method as a decomposition method rather than a clustering method.


Other methods for exploration and mining of networks include the work of Abello et al. \cite{abello02}. They introduced efficient network visualization techniques that allow users to identify useful facts about actors and relations through interactive navigation. They used abstraction levels starting from an overview to close-up views where the navigation is based on the notion of graph-slices. Henry et al.\cite{henry07} present the idea of NodeTrix, which is only a hybrid representation of node-link diagrams and adjacency matrix to better support the analysis of community structures. The representation coupled with various interaction techniques help users to create useful visualizations. Focus + Context techniques \cite{boutin06} are another way to reduce the node clutter for visual analysis and mining. These techniques allow users to simultaneously access overview information as well as focus on interesting groups. Kreuseler and Schumann\cite{kreuseler02} used an approach where they combine both analytical and visualization methods to interactive data exploration. They proposed a framework for visual data mining where the objective is to have a better understanding of the information space. 


\section{Data Sets}\label{sec::data}

For experimentation purposes, we selected three datasets. We have transformed these graphs into simple, undirected and unweighed graphs where required as the proposed method deals with simple graphs only. We only consider the biggest connected components for each dataset when the graphs were originally disconnected. 
\textbf{Geometry Network} is a collaboration network of authors where two authors are connected by an edge if they have co-authored an article together. The network contains 3621 nodes and 9461 edges (\url{http://vlado.fmf.uni-lj.si/pub/networks/data/}. \textbf{Dblp2008 Network} is another co-authorship network constructed from the DBLP database \url{http://www.informatik.uni-trier.de/~ley/db/} for the year 2008 which contains 93498 nodes and 260152 edges. \textbf{AirTransport Network} is a network of air traffic between cities~\cite{rozenblat08} having 1540 nodes and 16523 edges. \textbf{Opte Network} is an Internet Tomography network which is a collection of routing paths from a test host to other networks on the Internet. The database is available to the public from the Opte Project website (\url{http://opte.org/}). The network has 35836 nodes and 42387 edges. \textbf{Protein Network} is a Protein-Protein interactions network~\cite{gavin02}. The data is available from \url{http://dip.doe-mbi.ucla.edu/dip} and contains 1246 nodes and 3142 edges. 

\section{Topological Decomposition}\label{sec::topological}

In this section, we introduce the idea of \textit{Degree Induced Subgraphs} (DIS) which is a subgraph created by imposing constraints on node degrees. We define two such graphs:

\textbf{Definition 1:} \textbf{Max$_d$-DIS} is an induced subgraph of $G$ with vertex set $V'$ such that nodes in $V'$ have maximum degree $d$ in $G$. Mathematically for a graph $G(V,E)$ where $V$=nodes and $E$=edges, the Max$_d$-DIS is defined as an induced subgraph $G'(V',E')$ such that $V' \subseteq V$ and $\forall u \in V', deg_G(u)\leq d$ where $d$ can have values from $0$ to the maximum node degree possible for the network under consideration. 

\textbf{Definition 2:} \textbf{Min$_d$-DIS} is an induced subgraph of $G$ with vertex set $V'$ such that nodes in $V'$ have minimum degree $d$ in $G$. Mathematically for a graph $G(V,E)$ where $V$=nodes and $E$=edges, the $Min_d-DIS$ is defined as an induced subgraph $G'(V',E')$ such that $V' \subseteq V$ and $\forall u \in V', deg_G(u)\geq d$ where $d$ can have values from $0$ to the maximum node degree possible for the network under consideration. 


\subsection{Max$_d$-DIS: A closer look}\label{sec::maxddis}

\begin{figure}[t]
\begin{center}
\includegraphics[width=0.4\textwidth]{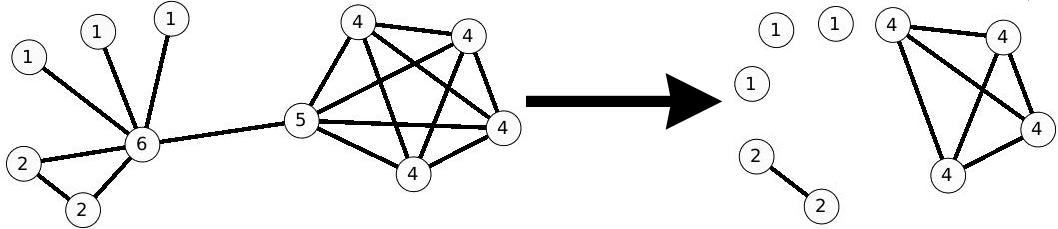}
\end{center}
\caption{An example of Max$_{d}-$DIS before and after calculating Max$_{4}-$DIS.}
\label{fig::egmaxddis}
\end{figure}

Consider a graph shown in Figure~\ref{fig::egmaxddis} (left) and we calculate the Max$_{4}$-DIS (right) in the figure. The nodes are labelled by their degree in the entire graph. Calculating the Max$_{4}$-DIS, the nodes with degree $5$ and $6$ are removed and an induced subgraph from the remaining nodes is obtained. In this example, it is quite clear that in Max$_{4}$-DIS, the nodes break into disconnected components. Thus a simple deduction is that the entire network was connected by high degree nodes, as soon as we removed them, the network broke into smaller connected components. Another interesting observation is that we can study how edges are distributed in low degree nodes as by definition, the Max$_d$-DIS considers only low degree nodes for low values of parameter $d$. An important focus in the study of scale free networks has been the role of \textit{hubs}~\cite{barabasi99} where nodes with very high degree stand out in networks. One way to look at the Max$_d$-DIS is that we want to avoid the hubs and look at how nodes connect to each other in their absence. 

\begin{figure}
\begin{center}
\includegraphics[width=0.31\textwidth]{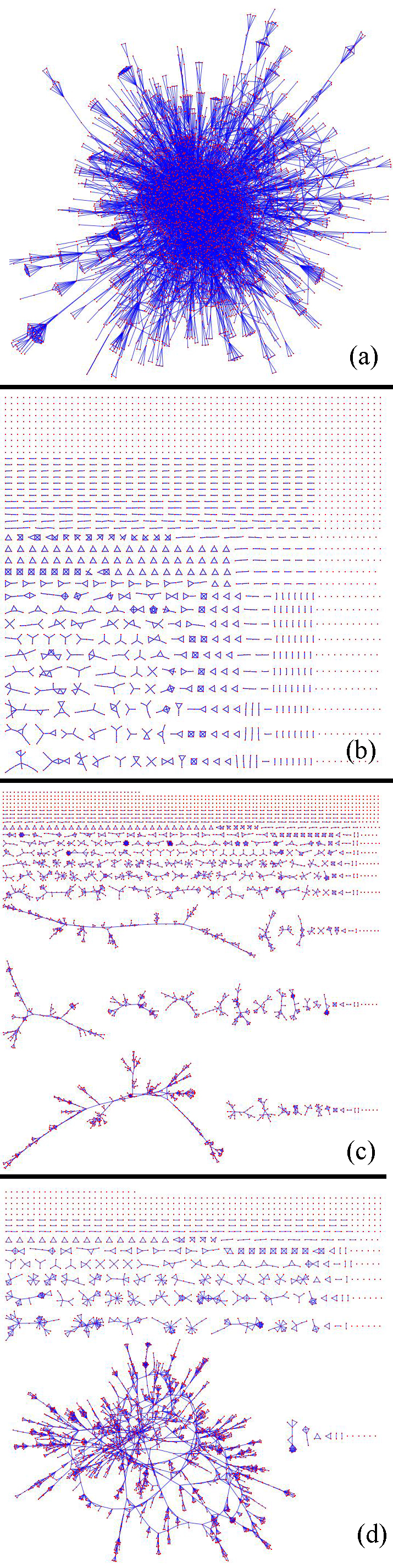}  
\end{center}
\caption{Visualization of Max$_d$-DIS graphs for the Geometry network. (a) Entire Network (b) Max$_{5}$-DIS (c) Max$_{10}$-DIS (d) Max$_{15}$-DIS.}
\label{fig::geomax}
\end{figure}

Lets take another example from a real world graph, that of the Geometry network by drawing several Max$_d$-DIS graphs. The graphs are drawn using a force directed algorith FMMM \cite{hachul05}. This algorithm puts nodes densely connected to each other closer in the layout and pushes nodes that are not connected to each other at distances. These algorithms are ideally suited for visual detection of community structures in networks. But in case of networks having scale free properties, where lots of nodes connect to only a few nodes, it becomes difficult to visually identify the presence of these communities. This is because nodes with very high connectivity are placed in the center of the layout and create an unreadable and tangled drawing in the center. Nodes of low degree are placed towards the outer periphery and are sparsely connected to each other. In Figure~\ref{fig::geomax}(a), we can see the entire network drawn using FMMM algorithm. The network contains over 3500 nodes which makes it quite difficult to see anything interesting in the network. 

Figure~\ref{fig::geomax}(b) shows the Max$_{5}$-DIS subgraph which contains $2757$ nodes which is more than $76\%$ of the total nodes in the entire graph. This suggests that most of the nodes in the network have low degree. Most of the time, our focus is towards the high degree nodes that stand out in the analysis of these complex networks. But the majority of the nodes have low node degree and we stress that analysis methods should also focus on these nodes as they are in majority and influence the overall behavior of the network to a large extent.

There are $1481$ edges in the subgraph which makes the average node degree to be $0.53$ as compared to the overall value of $2.6$ which is a huge difference in the context. The maximum node degree for the entire network is $102$. This means that most of these low degree nodes tend to connect with `higher' degree nodes. Note that we emphasize the relative degree and use `higher' instead of `highest', this is to suggest that since we are analyzing Max$_{5}$-DIS with degree limit $5$, these nodes might end up connecting to nodes with degrees $6,7,8$ and not necessarily with nodes of degree closer to the highest node degree which is $102$. We will have a look at this when we analyze the Max$_{10}$-DIS and Max$_{15}$-DIS graphs.

Another interesting observation is the number of connected components, $1537$ and there are lots of nodes with degree $0$ in the Max$_{5}$-DIS. Remember that the entire network is a single connected component, as high degree nodes appear, these smaller connected components merge to form one big single connected component. Finally, an important observation is about the structure formed by these low degree nodes when connecting to each other. We refer to the work of  Georg Simmel~\cite{simmel50} who introduced the concept of triads as a fundamental structure for social networks. The formation of triads has since been observed in many real world networks.

Figure~\ref{fig::geomax}(b) verifies the presence of triads as they are clearly visible. Triads are present not only in connected components of more than three nodes, but also in connected components of exactly three nodes. Moreover, we also observe cliques of sizes $4$ and $5$. By construction of this network, we expect cliques to be present. Since the data set is about articles in which it is common to have 3, 4 or more authors, it is quite obvious that we will find these cliques when visualizing the network. To avoid any false implications, we would like to refer to Figure~\ref{fig::egmaxddis} once again, notice the clique of $5$ nodes, when a Max$_d$-DIS is constructed, it no longer remains a clique of same size. Thus when analyzing a subgraph such as Figure~\ref{fig::geomax}(b), we should not conclude that there are only cliques of size $3$ or $4$, rather there are definitely cliques of larger sizes in the entire network.

Figure~\ref{fig::geomax}(c) shows the Max$_{10}$-DIS of the same network. Looking at the graph, we can immediately see that lots of small connected components from Figure~\ref{fig::geomax}(b) are now connected. Two such connected components are quite evident and there is another one which is of considerable size. In mathematics, this phenomena is called the emergence of a giant component~\cite{janson93}. The network changes drastically as certain links are introduced, and becomes a single connected component. In this case these links are introduced in the network by higher degree nodes that are responsible of connecting all these smaller components. The number of connected components in Max$_{10}$-DIS are $942$ as compared to $1537$ in Max$_{5}$-DIS, which is a considerable decrease as we do not have very high degree nodes.

\begin{figure}
\begin{center}
\includegraphics[width=0.48\textwidth]{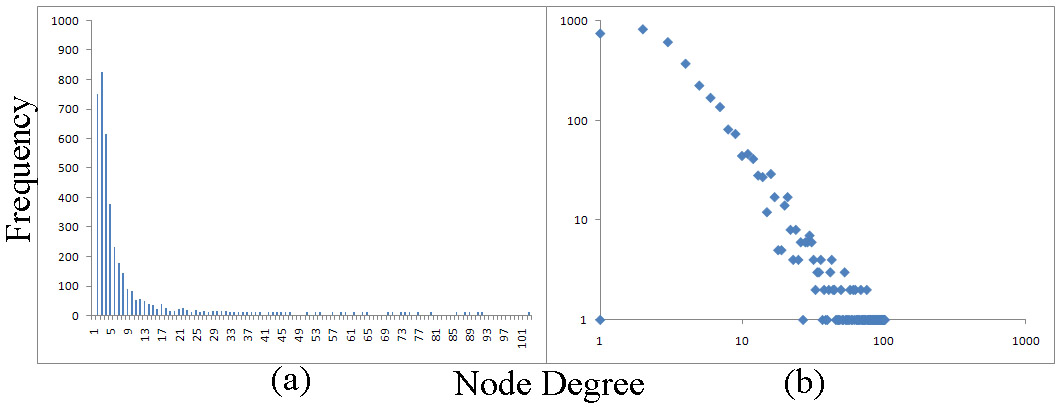}
\end{center}
\caption{(a)Histograms and (b) Log-Log scatter plot of the Degree Distribution of Geometry Network.}
\label{fig::deggeometry}
\end{figure}

Note that we still do not have the presences of very high degree nodes. If we look at the degree distribution of geometry network shown in Figure~\ref{fig::deggeometry}, we see that it is around degree value $10$ that the long tail starts to develop. In terms of clustering, this is quite significant. Consider if we want to group similar nodes, we can use this idea that in the absence of high degree nodes, this network breaks into smaller connected components, and from the subgraph, we can say that there are two big and a smaller cluster of nodes in the entire network. As high degree nodes are introduced in this network, the network becomes one big connected component and it becomes difficult to identify clusters. One observation from the degree distribution is that the highest frequency of nodes is that $60\%$ of the total nodes have degree less than $4$. 

And finally we move to Figure~\ref{fig::geomax}(d) which shows the Max$_{15}$-DIS. A clear development in this graph is the formation of the big connected component which comprises of $2133$ nodes and $3523$ edges. Notice that this graph only contains nodes with a maximum node degree of $15$ in the entire network as compared to the highest node degree of $102$. This shows that approx.\ $59\%$ of the nodes are linked in a single connected component without the presence of very high degree nodes. 

An obvious question arises, how we define `high' or `very high' degree? there is no concrete definition, but a rather vague estimation can be made by using heuristics. If we compare $15$ with $102$, the gap is quite wide and is easy to say that the value $15$ does not really represent `high' values of degree. On the other hand, if we compare $15$ with the average degree which is $2.6$, the value is quite high and can be considered as a high value for node degree. If we look in terms of percentage, the Max$_{15}$-DIS contains $3413$ nodes which is $94\%$ of the total nodes. If we consider the top $5\%$ nodes as the high degree nodes, Max$_{15}$-DIS suggests that in the absence of high degree nodes, $59\%$ of the nodes in the network are still connected. Thus we suggest that the Max$_{d}$-DIS can be used as a method to study an important feature of scale free networks studied by Albert \textit{et al.}~\cite{albert00} i.e.\ scale free networks are robust to random loss of nodes but fragile to targeted worst-case attacks on hubs.

\begin{figure}
\begin{center}
\includegraphics[width=0.32\textwidth]{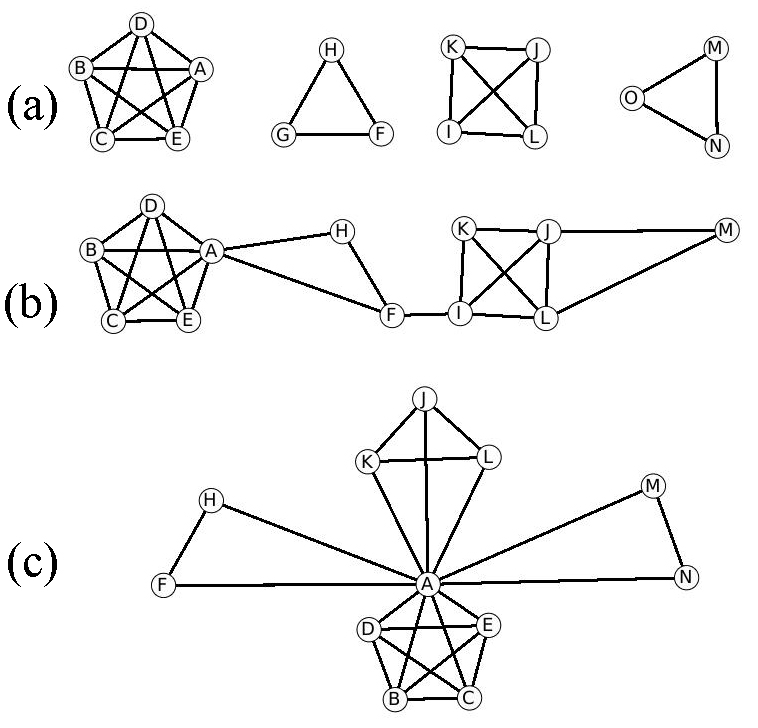}
\end{center}
\caption{(a)Four cliques representing different articles in a co-authorship network (b) The cliques are combined to form high average path length with certain nodes having higher degree (c) The cliques are combined to form low average path length with Node A standing out as a very high degree node.}
\label{fig::connectivityp}
\end{figure}

Looking at the average degree in this subgraph, for the $94\%$ of the nodes in the entire network is $1.33$ as compared to overall value of $2.6$ for the entire network. Thus the high degree nodes heavily influence the average degree of nodes in the entire network. Obviously this is a direct implication of the long tail in the degree distribution. The longer the tail, the bigger would be the difference in the average node degree of low degree nodes and the over all network. 

About the connectivity of the nodes and the three bigger connected components in Figure~\ref{fig::geomax}(c), there is an interesting observation about their average path lengths. The three components have values of $12.9$, $12.4$ and $9.7$ which when compared to the overall value of $5.31$ are quite high. Even the biggest connected component in Figure~\ref{fig::geomax}(d) has a very high average path length of $12.1$. As the high degree nodes appear in the network the average path length drops considerably in the network. If we look closely at Figure~\ref{fig::geomax}(a,b,c), we understand an important connectivity principle of these networks. When nodes connect to each other through `higher' degree nodes, the average path length is a bit higher as the connections form long paths, and when nodes connect through `very high' degree nodes, the average path length is lower. 

We further explain this using an example of co-authorship network. Consider the four cliques in Figure~\ref{fig::connectivityp}(a) representing four different articles, where the number of authors for the four cliques are 5,3,4,3 and each node is labelled with alphabets from A to O. The four cliques can be connected to each other by two principles. The first one is shown in Figure~\ref{fig::connectivityp}(b), which is in the absence of very high degree nodes. Consider a person authoring two articles with two different set of people, in that case, that node of that person will be between two cliques. From the example in the figure we consider person A and G are the same person and thus they connect the two cliques as shown in Figure~\ref{fig::connectivityp}(b). Another way to connect two cliques is to add edges between people writing different articles. In this example we add an edge between F and I, suggesting that earlier these people worked with a different set of people, but now they have collaborated to write together. Similarly, person J is the same as O, and L is the same person as N. This connectivity pattern introduced is due to the absence of very high degree nodes as, in this connected network, no node has a very high degree. The result is that we get a long string of cliques connected to each other just as we saw in Figure~\ref{fig::geomax}(c,d). On the other hand, another way to connect these cliques is shown in Figure~\ref{fig::connectivityp}(c) where a single person co-authors many articles with other people, in this case, that person is shown as node A, which collaborates with all other authors. This gives a certain node, a very high node degree and reduces the overall average path length to a large extent. Both these connectivity patterns are present in the geometry network as we see long paths in Figure~\ref{fig::connectivityp}(c,d) and short paths in Figure~\ref{fig::connectivityp}(a).

\begin{figure}
\begin{center}
\includegraphics[width=0.35\textwidth]{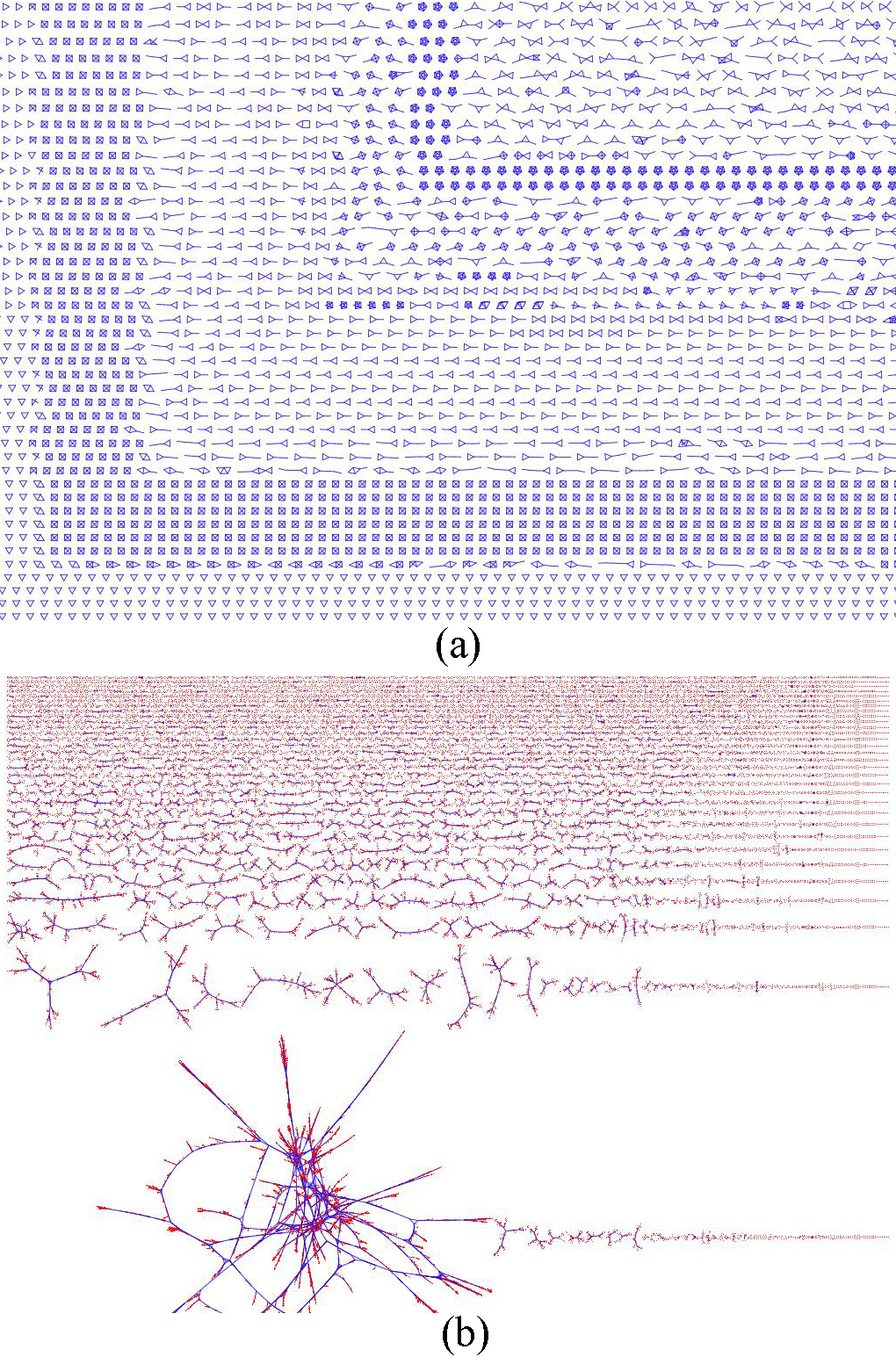}
\end{center}
\caption{Visualization of Max$_d$-DIS graphs for the Dblp2008 network. (a) Part of  Max$_{5}$-DIS (b) Part of Max$_{10}$-DIS.}
\label{fig::dblpmax}
\end{figure}

Lets summarize what we have explained in this section. We have tried to analyze the Geometry network using the Max$_d$-DIS decomposition. We identified several interesting observations such as:

\begin{enumerate}
	\item Studying how edges are distributed in low degree nodes
	\item Observe the structure of networks such as the formation of \textit{triads} and \textit{cliques} of bigger sizes	
	\item Analyze the connectivity of nodes in the absence of \textit{hubs} and see if a network is fragile to targeted attacks
	\item Break up of nodes in several disconnected components in the absence of high degree nodes, which motivates the idea of possibly grouping the connected components as clusters
	\item For higher values of $d$, the smaller components begin to merge into a single connected component even if very high degree nodes are not present
	\item The average path length of biggest connected components in the DIS subgraphs is considerably higher in the absence of very high degree nodes indicating that very high degree nodes are also responsible for reducing the overall average path length
	\item Two connectivity patterns are observed, one in the presence of higher degree nodes, and the other, when very high degree nodes appear. These patterns effect the average path length of a connected component 
\end{enumerate}

All of the above findings have been studied using other methods and in most cases, different methods to study different properties. Our contribution is not what we can find using the proposed method but towards a unified and non-complex method to help domain experts operate on their own and extract knowledge without the help of a computer science expert. From this detailed analysis of the Geometry network, we will have a look at some other real world networks and try to see if we observe similar findings in other networks. 

%
\begin{figure}
\begin{center}
\includegraphics[width=0.3\textwidth]{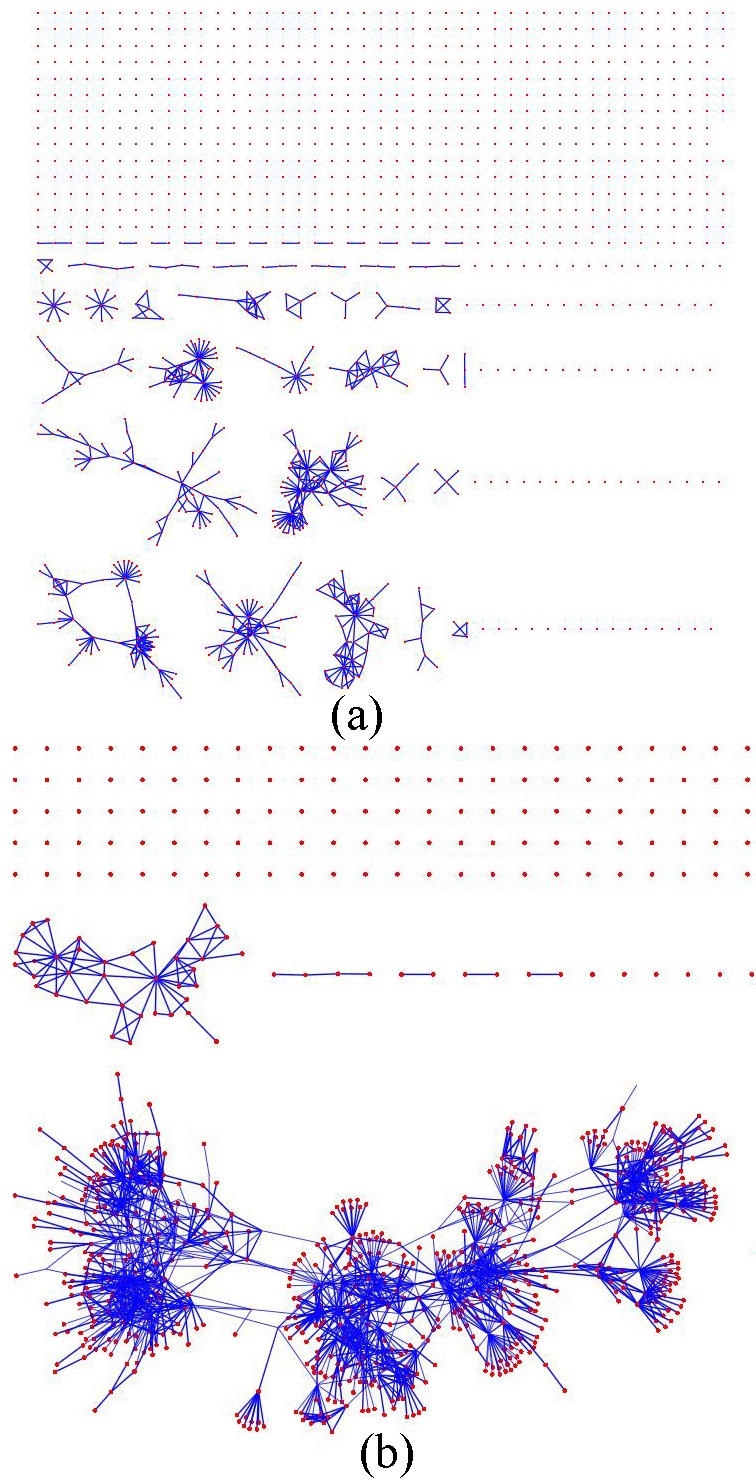}
\end{center}
\caption{Visualization of Max$_d$-DIS graphs for the AirTransport network. (a) Max$_{25}$-DIS (b) Part of Max$_{50}$-DIS.}
\label{fig::airmax}
\end{figure}
%


Next, we consider another co-authorship network, the Dblp2008 network. In Figure~\ref{fig::dblpmax}(a) and (b), we show the Max$_5$-DIS and Max$_{10}$-DIS of the network. In both these figures, we show only part of the entire subgraphs as Max$_5$-DIS contains over $60000$ nodes and Max$_{10}$-DIS over $80000$ nodes. The analysis is quite similar to that of the Geometry network as in the Max$_5$-DIS we see lots of smaller connected components forming cliques of sizes between $1$ and $5$. In Figure~\ref{fig::dblpmax}(b) we clearly see that the smaller components start to combine into one big connected component. This suggests that even at degree value $10$, smaller connected components begin to merge into a single connected component.

We move to another example, the AirTransport network. Figure~\ref{fig::airmax}(a) shows the Max$_{25}$-DIS and Figure~\ref{fig::airmax}(b) shows part of the Part of Max$_{50}$-DIS. We choose relatively higher values of $d$ in this example because there were not sufficient nodes or edges to observe anything interesting in lower values of $d$. From Figure~\ref{fig::airmax}(a), we can see that there are many disconnected components in the network as the previous examples. Recall, this signifies that these networks require higher degree nodes to connected to each other. In the  structure of the network, we see that there are triads as well as stars in Figure~\ref{fig::airmax}(a). If we look at the overall clustering coefficient, the presence of triads is confirmed from the relatively high value of $0.49$. If we look closely at the biggest connected component in Figure~\ref{fig::airmax}(b) with the Max$_{50}$-DIS, we see that there are many star like structures, i.e.\ many nodes with degree $1$ connect to a single node. This makes a very interesting example to study as we see that this network is a good mix of the two fundamental structures that we have identified, the cliques and the stars. Moreover, the average node degree in Max$_{50}$-DIS is $1.49$, which is very low as compared to the overall average degree of $10.7$. Along with the highest node degree of $487$, these values suggest that the high degree nodes are very well connected to other nodes in the network and thus push the average node degree to a high value of $10.7$. The biggest connected component in Figure~\ref{fig::airmax}(b) has $832$ nodes and $1941$ edges which is $54\%$ of the total nodes. The average path length of the nodes in this component is $5.9$ as compared to the overall value of $2.9$ and the considerable difference again suggests that very high degree nodes are responsible for the decrease in the overall average path length of the entire network.



\begin{figure}
\begin{center}
\includegraphics[width=0.42\textwidth]{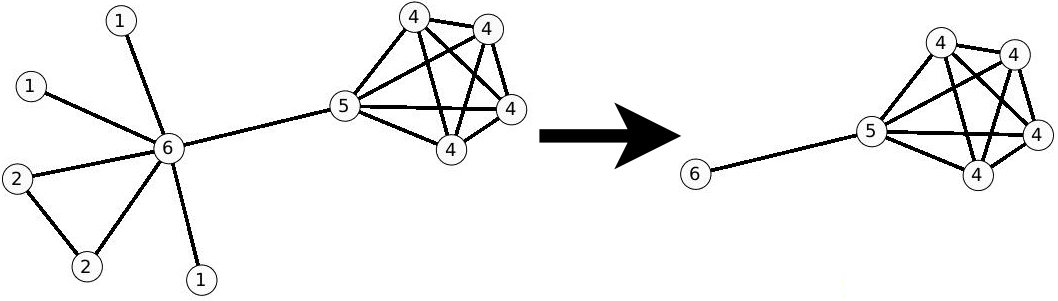}
\end{center}
\caption{An example of Min$_{d}-$DIS before and after calculating Min$_{4}-$DIS.}
\label{fig::egminddis}
\end{figure}

\subsection{Min$_d$-DIS: A closer look}\label{sec::minddis}

\begin{figure}
\begin{center}
\includegraphics[width=0.45\textwidth]{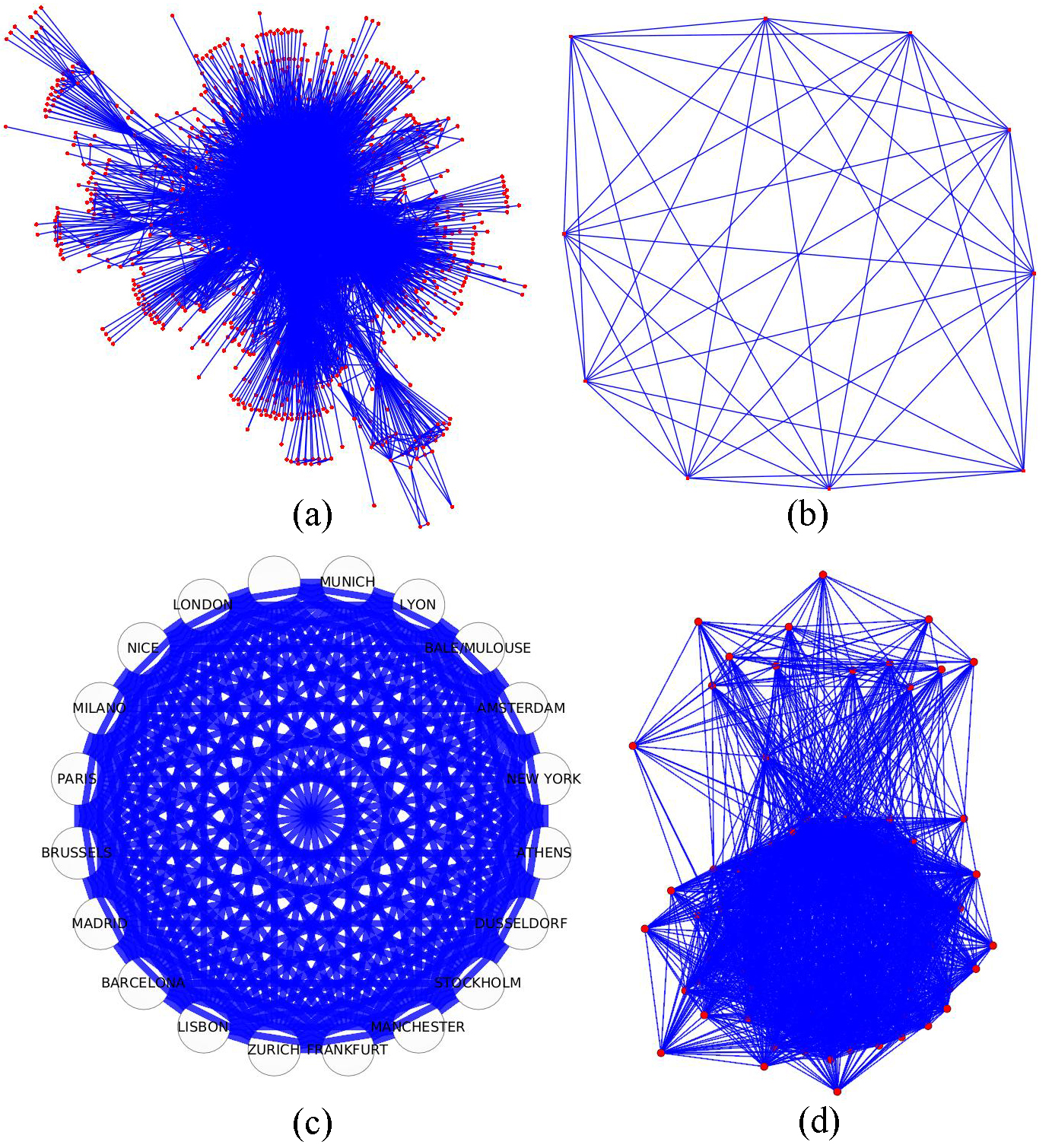}
\end{center}
\caption{Visualization of Min$_d$-DIS graphs for the AirTransport network (a) Entire Network (b) Min$_{250}$-DIS with the $10$ highest degree nodes (c) Min$_{185}$-DIS with the $20$ highest degree nodes (d) Min$_{94}$-DIS with the top $5\%$ highest degree nodes.}
\label{fig::airmin}
\end{figure}

In this section, we present the details of Min$_d$-DIS decomposition and how it can help us to analyze networks. Consider the graph in Figure~\ref{fig::egminddis} on the left and its Min$_4$-DIS on the right. Only the nodes with degree at least $5$ are left in the subgraph and the low degree nodes are removed. This method by definition considers only high degree nodes and thus helps us to have a look at how edges are distributed in high degree nodes of a network. In terms of analysis of real world networks, lets consider the AirTransport network for a detailed analysis using Min$_{d}-$DIS. Figure~\ref{fig::airmin}(a) shows the entire network containing $1540$ nodes and $16523$ edges. 

\begin{figure}
\begin{center}
\includegraphics[width=0.3\textwidth]{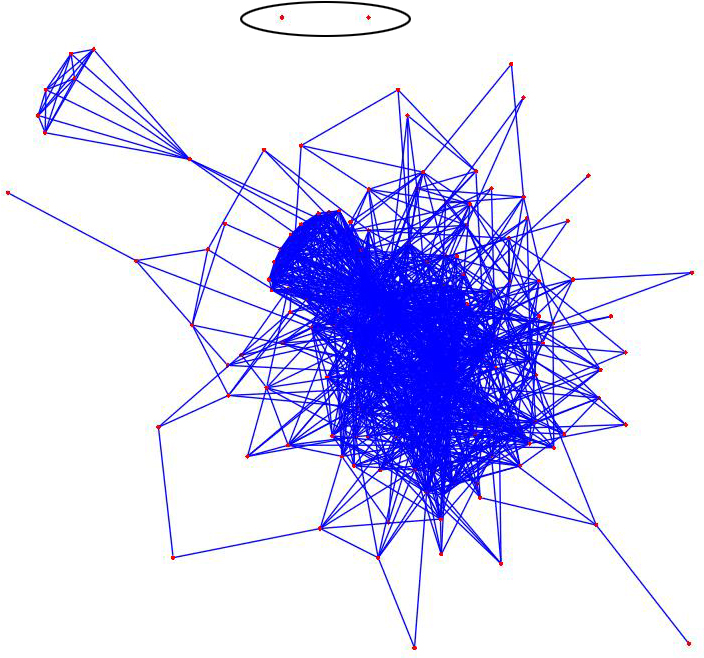}
\end{center}
\caption{Min$_{17}$-DIS of Geometry network with $5\%$ highest degree nodes.}
\label{fig::geomin5p}
\end{figure}

\begin{figure}
\begin{center}
\includegraphics[width=0.33\textwidth]{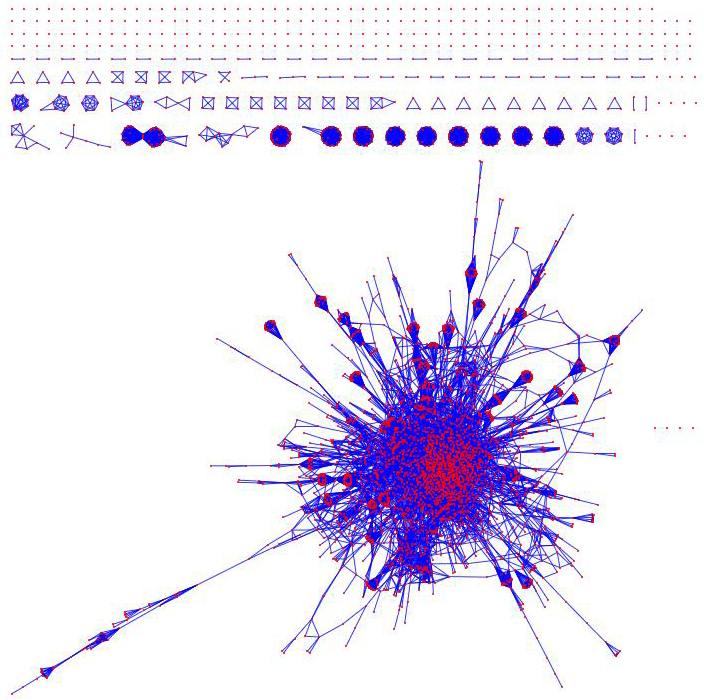}
\end{center}
\caption{Min$_{17}$-DIS of Dblp2008 network with $5\%$ highest degree nodes.}
\label{fig::dblpmin5p}
\end{figure}

%

Figure~\ref{fig::airmin}(b) shows the Min$_{250}-$DIS of the network showing the top $10$ highest degree nodes of the network connected through $44$ edges, which is just one less to make it a clique. This suggests that the worlds most widely connected airports have all a direct flight to each other with the exception of one case. The high degree of a city refers to its many different connections to other cities and not heavy traffic nor does it refer to the world's busiest airports. Figure~\ref{fig::airmin}(c) shows the top $20$ widely connected airports in the world drawn using a circular layout. We have also labelled the nodes with the city names. The nodes are connected through $189$ edges, missing the only edge for it to be a clique. This high connectivity of high degree nodes essentially comes from the design strategy of airlines and connectivity of any two cities of the world. Since the idea is to minimize the number of hops required to go from one place to the other, all the hubs in the network are connected directly to each other. Figure~\ref{fig::airmin}(d) shows the top $5\%$ high degree nodes of the network. There are $77$ nodes connected through $1822$ edges which is an average degree of $23$. In this subgraph, this high connectivity can be measured in terms of average path length, which is $1.3$ as compared to the average path length value for the entire graph which is $2.93$. This is a huge difference in the context. Note that there is not a single node disconnected in this network, thus high connectivity of hubs has been used as a method to increase the efficiency of transportation where the criteria is obviously minimizing the hops as described earlier.

This example shows an application of how Min$_d$-DIS can help analyze the connectivity of hubs in real world networks. From the above example, we learned that the hubs are very tightly connected to each other. We used the average path length to quantify how close the nodes are to each other. We summarize the type of analysis that can be performed using Min$_d$-DIS below:

\begin{itemize}
	\item Study the distribution of edges in high degree nodes, \textit{hubs}.
	\item Observe the connectivity pattern of high degree nodes and see how they connect to each other in different networks. We use Average path length to quantify how closely the hubs are connected to each other.	
\end{itemize}

Next, we consider the Geometry and Dblp2008 networks. For each of these networks, we took the Min$_d$-DIS with approximately $5\%$ highest degree nodes. The results are shown in Figures \ref{fig::geomin5p} and \ref{fig::dblpmin5p}  for these networks.

The Geometry network contains $179$ nodes and $1384$ edges. There are only two nodes that are not connected to the biggest connected component as shown in Figure~\ref{fig::geomin5p}. The average degree of this sub graph is $7.7$ as compared to $2.6$ of the overall network, which shows that these high degree nodes tend to connect to each other. The average path length of the big connected component is $2.4$ as compared to $5.3$ of the entire network, which again is a considerable difference. In this network, obviously the number of hops is not a criteria for efficiency, as it was for the AirTransport network. Here, the social contacts of a research community play an important role to keep the network well integrated where researchers collaborate to many different people, which results in a low average path length for this subgraph. We go back to the description of this network and the way this network is built. The data set was obtained from Computational Geometry Database which contains citation record of people working in this domain. The high connectivity of nodes within and the low average path length of these high degree nodes can be justified by the fact that people working in the same scientific domain have a higher probability of interacting with each other. We can compare the behavior of high average degree and low average path length of this network with that of AirTransport network as being similar, although the semantics and the reasoning behind this development are quite different. The other co-authorship network is the Dblp2008 which is significantly different from the Geometry network. Figure~\ref{fig::dblpmin5p} shows the Min$_{17}$-DIS of the network with $3872$ nodes and $20828$. There are $326$ number of connected components which is quite different from the previous two networks analyzed using Min$_d$-DIS. Although there is one big connected component, but there are many disconnected components that are themselves well connected to other nodes forming cliques. We call this behavior, local peaks as these are people working in different scientific domains and publishing many articles. Their domains do not cross necessarily and thus they are high publishing authors interacting with their restricted research community but not with people of other domains.



The Min$_d$-DIS reveals interesting structural behavior of real world networks which can be quite useful when analyzing and improving the efficiency of these networks. We found two interesting phenomena about the way high degree nodes can connect to each other.

\begin{itemize}
	\item Nodes with high degree are well connected to each other forming one big connected component with low average path lengths (comparing with respective average path lengths of their entire networks) \textit{OR}
	\item Nodes with high degree break into several connected components and have a path to each other through lower degree nodes
\end{itemize}


We studied the Max$_d$-DIS graphs for values $d=(5,10,15)$ and Min$_d$-DIS graphs for values top $5\%$ nodes. These values were chosen based on our analysis of different real world graphs. One quantitative method to select these values is based on the degree distribution and the long tail behavior. As discussed previously, the tangled layout of these graphs is due to the very high degree nodes, which can be detected by looking at the low frequency values or where the long tail approximately starts. This value need not be perfect as an approximation works well as shown in the case studies. To study how edges are distributed in low degree nodes (Max$_d$-DIS), values below the starting point of the tail can be used and to study how edges are distributed among high degree nodes (Min$_d$-DIS), values above this starting point can be used.

The time complexity of the overall method depends on the individual techniques applied at each step of the method, be it analysis or visualization. The decomposition itself is highly efficient and when slow metrics such as average path length is applied to small subgraphs, there is an exponential reduction in the time complexity making the overall method quite efficient. As for the scalability of the proposed method, we have studied graphs of around 100000 nodes and 250000 edges  (DBLP2008 network) and it took only less than 1 second to generate all possible Max$_d$-DIS and Min$_d$-DIS using the Tulip Software library on a standard Intel Core i3 machine without the visual representation. The visual representation depends on the layout algorithm used and FMMM is known to be one of the fastest layout algorithms producing readable layout drawings.

\section{Conclusions and Future Research Prospects}\label{sec::conclusions}

In this paper, we have introduced a method to analyze networks based on the topological decomposition and visualization of networks. The most important contribution of this method is that it enables us to visualize networks and see how nodes and edges connect to each other in different networks. Combining metrics and visualization technique to analyze complex large size networks proves to be quite useful as the method allows us to understand different connectivity behaviors as networks change from one domain to the other. 

We found some interesting behavior in the way the edges are distributed in high and low degree nodes. Based on this distribution, we can actually see that as a function of degree, dense subgraphs can be present in high degree and/or low degree nodes. Some interesting results about the structure of networks were discovered, or re-discovered such as the presence of triads, cliques of higher degree, star-like structures, showing the effectiveness of the proposed method. 


The topological decomposition proposed opens new dimensions in the field of visual data mining as complex networks can be simplified using the proposed method. Connectivity of nodes can be studied as a function of varying node degree of a network and helps us to discover how edges are distributed in a network. We see this method as an advancement towards the better understanding of complex networks and an important tool to analyze networks for various applications such as searching in networks, finding interesting patterns, studying the connectivity behavior of nodes in networks and develop local as well as group level metrics.

\bibliographystyle{IEEEtran}
\bibliography{visu}
%
%
%

\end{document}